\documentclass[aps,pra,superscriptaddress,preprint]{revtex4-1}
\usepackage{subfigure}
\usepackage{xcolor}
\usepackage{graphicx}
\usepackage{tikz}
\usepackage{bm}
\usetikzlibrary{shapes.geometric, arrows, positioning}
\usepackage{amsmath}
\usepackage{float}
\usepackage{ulem}

\usepackage[colorlinks,linkcolor=blue,anchorcolor=blue,citecolor=blue,urlcolor=blue]{hyperref}

\begin{document}

		\title{Anisotropic dispersion relation of ultralight Bose gases in modified Newtonian dynamics}
		\author{Ning Liu}
		\email{ningliu@mail.bnu.edu.cn}
		\address{School of Physics and Astronomy, Anqing Normal University, Anqing 246133, China}
		\address{Institute of Astronomy and Astrophysics, Anqing Normal University, Anqing 246133, China}
		\address{Key Laboratory of Multiscale Spin Physics (Ministry of Education), Beijing Normal University, Beijing 100875, China}
		
		\begin{abstract}
			We investigate the dispersion relation of collective modes in ultralight Bose gases under Modified Newtonian Dynamics (MOND). Starting from the coupled Gross–Pitaevskii and MOND Poisson equations, we derive an anisotropic dispersion relation that depends on the angle between the perturbation wavevector and the background gravitational field. This anisotropy arises directly from the nonlinear structure of the MOND field equation and leads to a direction-dependent Jeans instability, with critical wavelengths and masses varying with orientation. Our results provide a distinctive signature of MOND in a quantum astrophysical context and suggest that ultralight Bose gases can serve as novel probes of modified gravity.
		\end{abstract}

		\maketitle

	\section{Introduction}
	Self-gravitating Bose-Einstein condensates (SGBEC) are macroscopic quantum systems in which a Bose gas is bound by its own gravitational field. Described by a macroscopic wave function coupled to a gravitational potential, these systems bridge quantum many-body physics and astrophysics. Specially, SGBEC can form stable, self-bound quantum structures and have been studied in two main astrophysical contexts. The first context treats SGBEC as a candidate for dark matter, known as Bose-Einstein condensate dark matter~\cite{boehmer2007can, das2015dark, chavanis2017dissipative}. 
	In this framework, ultralight bosons condense into a superfluid phase, and the system is governed by the Gross--Pitaevskii--Poisson (GPP) equations~\cite{chavanis2011mass, chavanis2020jeans}. The GPP equations naturally produce a solitonic core surrounded by an incoherent halo, offering a compelling resolution to the small-scale crises of the standard cold dark matter model, such as the core-cusp problem~\cite{chavanis2025review, paredes2020optics}. The second context treats SGBEC as independent astrophysical objects, often termed boson stars or axion stars~\cite{kaup1968klein,ruffini1969systems,seidel1990dynamical,jetzer1992boson}. 
	Such objects may form through gravitational cooling of scalar fields or early-universe phase transitions, and their existence does not depend on whether they account for the entirety of dark matter. 
	
	SGBEC have been studied extensively in Newtonian gravity and general relativity, revealing rich equilibrium and stability properties~\cite{braaten2019colloquium,visinelli2021boson,liebling2023dynamical}. Recent advances have further revealed the dynamics of these quantum structures, including the stability and collisions of vortex structures~\cite{nikolaieva2023stable, asakawa2024corotation}, the onset of quantum turbulence during mergers~\cite{sivakumar2025revealing}, and self-similar growth~\cite{dmitriev2024self}. 
	Notably, in almost all of the aforementioned studies, the gravitational interaction is treated within Newtonian gravity (or general relativity for compact relativistic configurations). 
	
	However, the internal gravitational acceleration of a typical self-gravitating Bose gas composed of ultralight bosons is extremely feeble. For a non-interacting boson star, the mass-radius relation \(R \sim \hbar^2/(G M m^2)\) implies a surface gravity \(g = GM/R^2 \sim 10^{-10}\,\mathrm{m/s}^2\) for \(m \sim 10^{-22}\,\mathrm{eV}/c^2\) and \(M \sim 10^{11}M_\odot\). The straightforward scaling analysis reveals that the characteristic surface gravity of such a solitonic core is of order \(a_0\), the critical acceleration scale of Modified Newtonian Dynamics (MOND)~\cite{famaey2012modified}.

	MOND, proposed by Milgrom in 1983~\cite{milgrom1983modification}, postulates a departure from standard Newtonian gravity at accelerations below \(a_0 \approx 1.2\times10^{-10}\,\mathrm{m/s^2}\) to explain the observed flat rotation curves of galaxies without invoking particle dark matter. MOND posits that for accelerations below \(a_0\), the effective gravitational acceleration \(g\) relates to the Newtonian acceleration \(g_N\) via \(g = g_N / \mu(g/a_0)\), where \(\mu(x)\) is an interpolation function satisfying \(\mu(x) \to 1\) for \(x \gg 1\) (Newtonian regime) and \(\mu(x) \to x\) for \(x \ll 1\) (deep-MOND regime). This modification leads to a number of striking predictions, including flat rotation curves without dark matter and a baryonic Tully-Fisher relation \(M \propto v^4\)~\cite{milgrom1983modification}. The deep-MOND regime also exhibits scale invariance, a property absent in Newtonian gravity~\cite{milgrom2014mond}. In the non-relativistic limit, MOND is captured by the modified Poisson equation \(\nabla\cdot[\mu(|\nabla\Phi|/a_0)\nabla\Phi] = 4\pi G\rho\), which reduces to the standard Poisson equation when \(|\nabla\Phi| \gg a_0\). In principle, MOND can be interpreted either as a modification of Newtonian gravity or, more profoundly, as a modification of Newtonian mechanics in general. The latter, stronger interpretation has been ruled out experimentally~\cite{gundlach2007laboratory}. The version of MOND that remains under serious consideration is the one in which only the Lagrangian of the gravitational field is modified. We therefore adopt this widely studied gravitational formulation of MOND.
	
	A natural concern is that boson stars are often studied as dark matter candidates, whereas MOND avoids dark matter. The key point, however, is that boson stars can exist independently of the dark matter hypothesis—as compact objects formed from ordinary or exotic matter (e.g., axion clumps or gravitational cooling remnants). They are not invoked to explain galactic rotation curves. Instead, their collective excitations serve as probes of the underlying gravitational law, including its MOND modification. Thus, we use boson stars as probes of MOND, not as dark matter.
	
	In this paper, we investigate the dispersion relation of a uniform self-gravitating Bose gas governed by the coupled Gross--Pitaevskii and MOND Poisson equations. 
	Using the Bogoliubov--de Gennes formalism, we obtain an anisotropic dispersion relation that depends explicitly on the angle between the perturbation wavevector and the local background gravitational field. This anisotropy is a direct consequence of the nonlinear structure of the MOND field equation and vanishes in the Newtonian limit. Consequently, the Jeans instability criterion becomes direction-dependent, leading to anisotropic collapse scales that differ markedly from the standard Newtonian case. 
	
	The paper is structured as follows. Section~\ref{sec:model} presents the coupled GP-MOND model and its linearization. Section~\ref{sec:bdg} details the Bogoliubov--de Gennes analysis yielding the anisotropic dispersion relation. Section~\ref{sec:jeans} analyzes the direction-dependent Jeans instability and discusses astrophysical implications and future directions. We conclude in Section~\ref{sec:conclusion}.
	
	\section{Model}\label{sec:model}
	Before presenting the coupled GP-MOND equations, we first address a natural question: can the Gross--Pitaevskii equation, originally derived for laboratory Bose-Einstein condensates, legitimately describe a self-gravitating Bose gas on astrophysical scales? A large body of literature has established that under specific conditions, the answer is affirmative. Chavanis has systematically shown that the GPP system provides a consistent mean-field description of self-gravitating Bose gases, applicable both to boson stars and to dark matter halos when the boson mass is sufficiently small~\cite{chavanis2025review}. The key insight is that for ultralight bosons with masses on the order of $10^{-22}\,\mathrm{eV}/c^2$, the de Broglie wavelength becomes comparable to kiloparsec scales, making quantum coherence manifest on astronomical distances. Simultaneously, the diluteness condition $n a_s^3 \ll 1$ is satisfied over a wide range of densities characteristic of such systems, and the condensation temperature far exceeds the actual temperature of the cosmic gas, ensuring that the system is deep in the condensed phase. Under these combined conditions of macroscopic quantum coherence, diluteness, and effective zero temperature, the Gross--Pitaevskii equation genuinely captures the essential competition between quantum pressure, contact interactions, and self-gravity. The GPP framework has therefore been extensively adopted as a benchmark model for Bose stars, a standing that we inherit while extending the gravitational sector to incorporate MOND.
	
	The system consists of a self-gravitating BEC described by a macroscopic wave function \(\psi(\mathbf{r},t)\) and a gravitational potential \(\Phi(\mathbf{r},t)\). Its dynamics follows from the Lagrangian density
	\begin{equation}
		\mathcal{L}[\psi,\psi^*,\Phi] = \mathcal{L}_{\mathrm{GP}}[\psi,\psi^*] + \mathcal{L}_{\mathrm{MOND}}[\psi,\psi^*,\Phi],
		\label{eq:total_lagrangian}
	\end{equation}
	with the Gross–Pitaevskii (GP) part~\cite{pitaevskii2016bose}
	\begin{equation}
		\begin{aligned}
			\mathcal{L}_{\mathrm{GP}} &= \frac{i\hbar}{2}\left(\psi^*\partial_t\psi - \psi\partial_t\psi^*\right) 
			- \frac{\hbar^2}{2m}|\nabla\psi|^2 \\
			&\quad - V_{\mathrm{ext}}|\psi|^2
			- \frac{g}{2}|\psi|^4,
		\end{aligned}
		\label{eq:L_BEC}
	\end{equation}
	and the MOND gravitational part~\cite{bekenstein1984does}
	\begin{equation}
		\mathcal{L}_{\mathrm{MOND}} = -m\Phi|\psi|^2 
		- \frac{a_0^2}{8\pi G}\,
		\mathcal{F}\!\left(\frac{|\nabla\Phi|^2}{a_0^2}\right).
		\label{eq:L_MOND_int}
	\end{equation}
	Here \(m\) is the atomic mass, \(V_{\mathrm{ext}}\) an external trapping potential, and \(g = 4\pi\hbar^2 a_s/m\) the contact interaction strength, where \(a_s\) is the \(s\)-wave scattering length. The MOND acceleration scale is denoted by \(a_0\). The function \(\mathcal{F}(x^2)\) is related to the MOND interpolation function \(\mu(x)\) via \(\mu(x) = \mathcal{F}'(x^2)\). In the Newtonian limit \(|\nabla\Phi| \gg a_0\), one has \(\mu \to 1\), while in the deep-MOND limit \(|\nabla\Phi| \ll a_0\), \(\mu(x) \approx x\).
	
	Varying the action \(S = \int dt \int d^3\mathbf{r} \, \mathcal{L}\) with respect to \(\psi^*\) and \(\Phi\) yields the equations of motion. Variation with respect to \(\psi^*\) gives the GP equation
	\begin{equation}
		i\hbar\frac{\partial\psi}{\partial t} = 
		\left[-\frac{\hbar^2}{2m}\nabla^2 + V_{\mathrm{ext}} + g|\psi|^2 + m\Phi\right]\psi,
		\label{eq:GP}
	\end{equation}
	and variation with respect to \(\Phi\) leads to the MOND Poisson equation
	\begin{equation}
		\nabla\cdot\left[\,\mu\!\left(\frac{|\nabla\Phi|}{a_0}\right)\nabla\Phi\,\right] = 4\pi G\,m\,|\psi|^2.
		\label{eq:MOND_Poisson}
	\end{equation}
	Setting \(\mu = 1\) recovers the standard Poisson equation \(\nabla^2\Phi = 4\pi G m|\psi|^2\); Eqs.~\eqref{eq:GP} and \eqref{eq:MOND_Poisson} then reduce to the Gross–Pitaevskii–Poisson (or GP-Newton) system describing a self-gravitating BEC in Newtonian gravity.
	
	\section{Bogoliubov-de Gennes Theory} \label{sec:bdg}
	To study small-amplitude collective modes, we expand the wave function and gravitational potential around their equilibrium values:
	\begin{align}
		\psi(\mathbf{r},t) &= e^{-i\mu_c t/\hbar}\left[\psi_0 + \delta\varphi(\mathbf{r},t)\right], \\
		\Phi(\mathbf{r},t) &= \Phi_0(\mathbf{r}) + \delta\Phi(\mathbf{r},t),
	\end{align}
	with \(|\delta\varphi|\ll|\psi_0|\) and \(|\delta\Phi|\ll|\Phi_0|\). Here \(\psi_0\) is the equilibrium condensate wave function, \(\Phi_0\) the equilibrium gravitational potential, and \(\mu_c\) the chemical potential. Substituting these expansions into Eqs.~\eqref{eq:GP} and \eqref{eq:MOND_Poisson} and linearizing in the perturbations yields
	\begin{equation}
		i\hbar\frac{\partial\delta\varphi}{\partial t} = -\frac{\hbar^2}{2m}\nabla^2\delta\varphi + g n_0(\delta\varphi+\delta\varphi^*) + m\psi_0\delta\Phi,
		\label{eq:lin_GP}
	\end{equation}
	and
	\begin{equation}
		\nabla\cdot\left[\mu_0\nabla\delta\Phi + \frac{\mu_0'}{a_0}(\hat{e}\cdot\nabla\delta\Phi)\mathbf{g}_0\right] = 4\pi G m\psi_0(\delta\varphi+\delta\varphi^*).
		\label{eq:lin_Poisson}
	\end{equation}
	The equilibrium density is \(n_0 = |\psi_0|^2\). The background gravitational field \(\mathbf{g}_0 = -\nabla\Phi_0\) has magnitude \(\mathcal{G}_0 = |\mathbf{g}_0|\) and direction \(\hat{e} = \mathbf{g}_0/\mathcal{G}_0\). The coefficients \(\mu_0 = \mu(\mathcal{G}_0/a_0)\) and \(\mu_0' = d\mu/dx|_{x=\mathcal{G}_0/a_0}\) are the MOND interpolation function and its derivative evaluated at the scaled field strength.
	
	We consider a locally homogeneous background and look for plane-wave perturbations
	\begin{align}
		\delta\varphi(\mathbf{r},t) &= u e^{i(\mathbf{k}\cdot\mathbf{r}-\omega t)} + v^* e^{-i(\mathbf{k}\cdot\mathbf{r}-\omega t)}, \\
		\delta\Phi(\mathbf{r},t) &= w e^{i(\mathbf{k}\cdot\mathbf{r}-\omega t)} + w^* e^{-i(\mathbf{k}\cdot\mathbf{r}-\omega t)},
	\end{align}
	with constant amplitudes \(u\), \(v\), \(w\), wavevector \(\mathbf{k}\), and frequency \(\omega\). Substituting these forms into the linearized equations and matching Fourier coefficients leads to the Bogoliubov–de Gennes (BdG) equations~\cite{pethick2008bose}
	\begin{align}
		\hbar\omega u &= \frac{\hbar^2 k^2}{2m}u + g n_0(u+v) + m\psi_0 w, \\
		-\hbar\omega v &= \frac{\hbar^2 k^2}{2m}v + g n_0(u+v) + m\psi_0 w,
	\end{align}
	together with the Fourier-space MOND equation
	\begin{equation}
		\left[-\mu_0 k^2 - \frac{\mu_0'\mathcal{G}_0}{a_0}(\mathbf{k}\cdot\hat{e})^2\right] w = 4\pi G m\psi_0 (u+v).
		\label{eq:poisson_fourier}
	\end{equation}
	Defining the anisotropic denominator
	\begin{equation}
		D(\mathbf{k},\theta) = \mu_0 k^2 + \frac{\mu_0'\mathcal{G}_0}{a_0}(\mathbf{k}\cdot\hat{e})^2,
		\label{eq:Dk}
	\end{equation}
	we solve Eq.~\eqref{eq:poisson_fourier} for \(w\):
	\begin{equation}
		w = -\frac{4\pi G m\psi_0}{D(\mathbf{k},\theta)}(u+v).
		\label{eq:w_solution}
	\end{equation}
	Eliminating \(w\) from the BdG equations gives a \(2\times2\) eigenvalue problem. Its solvability condition yields the dispersion relation
	\begin{equation}
		\omega^2 = \frac{\hbar^2 k^4}{4m^2} + \frac{k^2}{m}\left[g n_0 - \frac{4\pi G m^2 n_0}{D(\mathbf{k},\theta)}\right],
		\label{eq:dispersion_general}
	\end{equation}
	which is the central result of our analysis. The factor \(D(\mathbf{k},\theta)\) depends on \((\mathbf{k}\cdot\hat{e})^2\), introducing an anisotropy: the excitation frequency varies with the angle \(\theta\) between \(\mathbf{k}\) and the background field direction \(\hat{e}\).
	
	In the Newtonian limit (\(\mathcal{G}_0 \gg a_0\)), \(\mu_0 \to 1\) and \(\mu_0' \to 0\), so \(D(\mathbf{k},\theta) \to k^2\) and Eq.~\eqref{eq:dispersion_general} reduces to the isotropic quantum Jeans dispersion
	\begin{equation}
		\omega^2 = \frac{\hbar^2 k^4}{4m^2} + \frac{g n_0}{m}k^2 - 4\pi G m n_0.
		\label{eq:dispersion_newton}
	\end{equation}
	
	In the deep-MOND regime (\(\mathcal{G}_0 \ll a_0\)), the interpolation function behaves as \(\mu(x) \approx x\), giving \(\mu_0 \approx \mathcal{G}_0/a_0\) and \(\mu_0' \approx 1\). Then
	\begin{equation}
		D(\mathbf{k},\theta) \approx \frac{\mathcal{G}_0}{a_0}\left[k^2 + (\mathbf{k}\cdot\hat{e})^2\right] = \frac{\mathcal{G}_0}{a_0}k^2(1+\cos^2\theta),
		\label{eq:Dk_deepMOND}
	\end{equation}
	with \(\cos\theta = \hat{k}\cdot\hat{e}\). Substituting into Eq.~\eqref{eq:dispersion_general} gives the deep-MOND dispersion relation
	\begin{equation}
		\omega^2 = \frac{\hbar^2 k^4}{4m^2} + \frac{g n_0}{m}k^2 - \frac{4\pi G m n_0 a_0}{\mathcal{G}_0(1+\cos^2\theta)}.
		\label{eq:dispersion_deepMOND}
	\end{equation}
	The anisotropy is now explicit: the gravitational term depends on \(\theta\). For \(\theta=0\) (wavevector parallel to \(\hat{e}\)), the denominator is \(2\mathcal{G}_0\), yielding the weakest gravitational correction; for \(\theta=\pi/2\) (perpendicular case), the denominator is \(\mathcal{G}_0\), giving a correction twice as strong.
	
	In the limit of negligible gravity (\(G \to 0\)), Eq.~\eqref{eq:dispersion_deepMOND} reduces to the familiar Bogoliubov excitation spectrum \(\omega^2 = \hbar^2 k^4/(4m^2) + (g n_0/m)k^2\). In the classical limit (\(\hbar \to 0\)), the quantum pressure term vanishes, giving the purely classical dispersion relation
	\begin{equation}
		\omega^2 = \frac{g n_0}{m}k^2 - \frac{4\pi G m n_0 a_0}{\mathcal{G}_0(1+\cos^2\theta)}.
		\label{eq:classical_limit}
	\end{equation}
	This classical expression highlights the anisotropic gravitational contribution without quantum effects.

	To highlight the scaling, we introduce the healing length \(\xi = \hbar/\sqrt{2 m g n_0}\) and the dimensionless quantities
	\begin{equation}
		\tilde{k}=k\xi,\qquad \tilde{\omega}=\frac{\hbar\omega}{g n_0},\qquad 
		\chi=\frac{4\pi G m^2\xi^2}{g},\qquad 
		\eta=\frac{\mathcal{G}_0}{a_0}.
	\end{equation}
	Here \(\chi\) measures the relative strength of gravity, and \(\eta \ll 1\) in the deep-MOND regime. Eq.~\eqref{eq:dispersion_deepMOND} then becomes
	\begin{equation}
		\tilde{\omega}^2 = \tilde{k}^4 + 2\tilde{k}^2 - \frac{2\chi}{\eta(1+\cos^2\theta)}.
		\label{eq:dispersion_dimensionless}
	\end{equation}
	
\begin{figure}[ht!]
		\centering
		\includegraphics[width=0.6\textwidth]{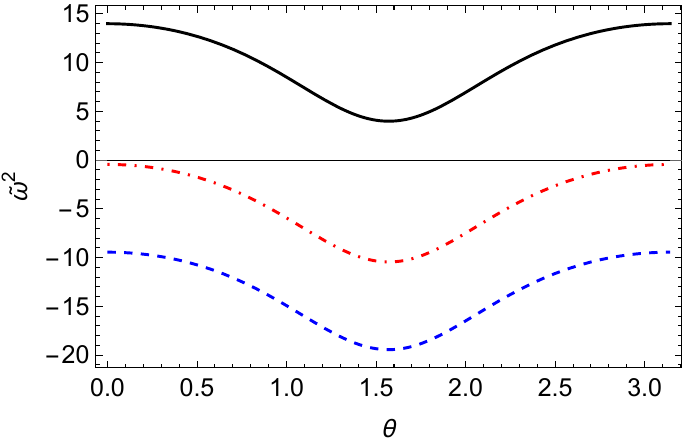}
		\caption{Squared frequency \(\tilde{\omega}^2\) versus angle \(\theta\) for fixed wavenumbers \(\tilde{k} = 0.5\) (dashed), \(\tilde{k} = 1.0\) (dot-dashed), and \(\tilde{k} = 1.5\) (solid), with \(\chi = 1\) and \(\eta = 0.1\).}
		\label{fig:omega_theta}
	\end{figure}

	To illustrate the anisotropy, we first plot in Fig.~\ref{fig:omega_theta} the dimensionless squared frequency \(\tilde{\omega}^2\) as a function of the angle \(\theta\) for several fixed dimensionless wavenumbers \(\tilde{k}\). For each \(\tilde{k}\), \(\tilde{\omega}^2\) decreases monotonically from \(\theta = 0\) to \(\theta = \pi/2\), confirming that perturbations perpendicular to the background gravitational field are more unstable—i.e., have a lower or more negative squared frequency—than parallel ones at the same wavenumber. Consequently, the critical wavenumber for instability (where \(\tilde{\omega}^2 = 0\)) is smaller for perpendicular perturbations and larger for parallel ones. This means that, to become unstable, modes parallel to the background field require a shorter wavelength (larger \(k\)) compared to perpendicular modes.
	
\begin{figure}[ht!]
		\centering
		\includegraphics[width=0.6\textwidth]{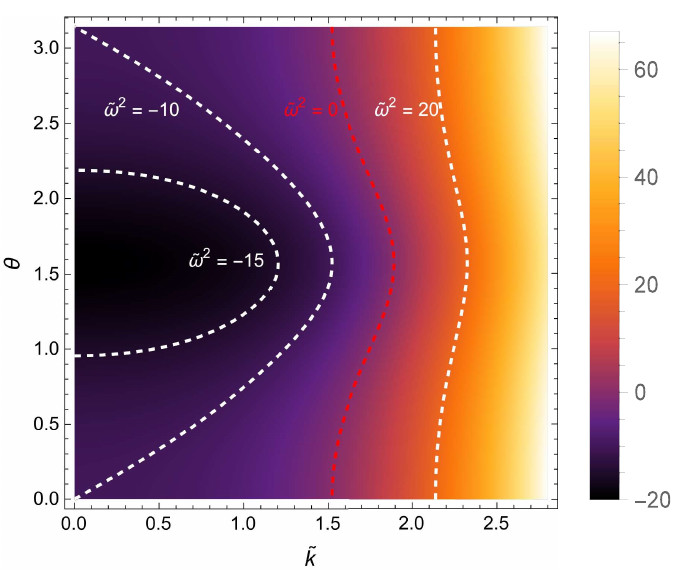}
		\caption{Contour plot of the squared frequency \(\tilde{\omega}^2\) in the \(\tilde{k}\)-\(\theta\) plane for \(\chi = 1\), \(\eta = 0.1\). The red dashed line indicates the stability boundary \(\tilde{\omega}^2 = 0\), separating the stable (right) and unstable (left) regions.}
		\label{fig:dispersion}
	\end{figure}

	The full anisotropic landscape is displayed in Fig.~\ref{fig:dispersion} as a contour plot of \(\tilde{\omega}^2\) in the \(\tilde{k}\)-\(\theta\) plane. The red dashed contour marks the stability boundary \(\tilde{\omega}^2 = 0\), which separates the parameter space into two regions: the stable collective-excitation region on the right (\(\tilde{\omega}^2 > 0\)) and the unstable, collapse-prone region on the left (\(\tilde{\omega}^2 < 0\)). The shape of this boundary clearly shows that the critical wavenumber for stability decreases as \(\theta\) increases from \(\theta = 0\) to \(\theta = \pi/2\), confirming the directional dependence of the Jeans instability.

	\section{Jeans Instability Analysis}\label{sec:jeans}
	In the long-wavelength limit $\tilde{k}\to0$, Eq.~\eqref{eq:dispersion_dimensionless} gives
	\begin{equation}
		\lim_{\tilde{k}\to0}\tilde{\omega}^2 = -\frac{2\chi}{\eta(1+\cos^2\theta)} < 0,
	\end{equation}
	indicating Jeans instability for all directions, with a growth rate that depends on $\theta$. The critical wavenumber $k_J$ at which $\omega=0$ satisfies
	\begin{equation}
		\tilde{k}_J^4 + 2\tilde{k}_J^2 = \frac{2\chi}{\eta(1+\cos^2\theta)},
	\end{equation}
	where $\tilde{k}_J = k_J\xi$ is the dimensionless critical wavenumber. Solving this quadratic equation for $\tilde{k}_J^2$ yields
	\begin{equation}
		\tilde{k}_J^2 = \sqrt{1 + \frac{2\chi}{\eta(1+\cos^2\theta)}} - 1,
	\end{equation}
	or equivalently
	\begin{equation}
		k_J(\theta) = \frac{1}{\xi}\sqrt{\sqrt{1 + \frac{2\chi}{\eta(1+\cos^2\theta)}} - 1}.
		\label{eq:k_J}
	\end{equation}
	This is a decreasing function of $\cos^2\theta$; hence $k_J$ is largest for $\theta=\pi/2$ and smallest for $\theta=0$.
	
	The Jeans wavelength, defined as $\lambda_J = 2\pi/k_J$, becomes
	\begin{equation}
		\lambda_J(\theta) = 2\pi\xi\left[\sqrt{1 + \frac{2\chi}{\eta(1+\cos^2\theta)}} - 1\right]^{-1/2},
		\label{eq:lambda_J}
	\end{equation}
	which is smallest for $\theta=\pi/2$ and largest for $\theta=0$. This anisotropic Jeans criterion contrasts sharply with the isotropic one in Newtonian gravity.
	
	The Jeans mass $M_J$, defined as the mass contained within a sphere of diameter $\lambda_J$, is given by
	\begin{equation}
		M_J(\theta) = \frac{4\pi}{3}\rho_0\left(\frac{\lambda_J(\theta)}{2}\right)^3 = \frac{\pi}{6}\rho_0\lambda_J^3(\theta),
	\end{equation}
	where $\rho_0 = mn_0$ is the mass density. Substituting Eq.~\eqref{eq:lambda_J} gives
	\begin{equation}
		M_J(\theta) = \frac{\pi}{6}\rho_0(2\pi\xi)^3\left[\sqrt{1 + \frac{2\chi}{\eta(1+\cos^2\theta)}} - 1\right]^{-3/2}.
		\label{eq:M_J}
	\end{equation}
	
	In the classical limit ($\hbar\to0$), the quantum pressure term vanishes and the dispersion relation reduces to Eq.~\eqref{eq:classical_limit}. Setting $\omega=0$ in that equation gives the classical Jeans wavenumber
	\begin{equation}
		k_{J,\mathrm{class}}(\theta) = \sqrt{\frac{4\pi G m^2 a_0}{g\,\mathcal{G}_0\,(1+\cos^2\theta)}},
	\end{equation}
	which likewise exhibits a clear $\theta$-dependence. Hence, the anisotropy of the Jeans scale is generic consequence of the MOND nonlinearity, persisting in both quantum and classical regimes.
	
\begin{figure}[htbp]
		\centering
		\includegraphics[width=0.6\textwidth]{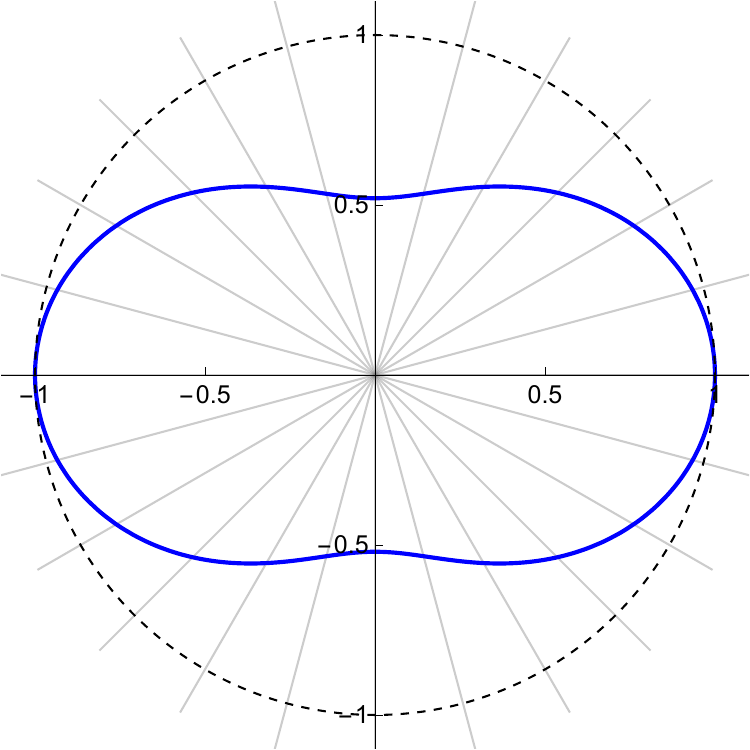}
		\caption{Polar plot of the normalized Jeans mass $\widetilde{M}_J(\theta)$. The radial coordinate is $\widetilde{M}_J(\theta)$; the dashed circle indicates the isotropic Newtonian limit (normalized to $1$). Parameters are the same as in Fig.~\ref{fig:omega_theta}.}
		\label{fig:mass_angular}
	\end{figure}

	To illustrate the angular anisotropy of the collapse scale, we introduce the normalized Jeans mass $\widetilde{M}_J(\theta) = M_J(\theta)/M_J(0)$, which measures the mass ratio relative to the parallel direction ($\theta=0$). From Eq.~\eqref{eq:M_J}, we have
	\begin{equation}
		\widetilde{M}_J(\theta)= \left[\frac{\sqrt{1 + \chi/\eta} - 1}{\sqrt{1 + \frac{2\chi}{\eta(1+\cos^2\theta)}} - 1}\right]^{3/2}.
	\end{equation}
	This function decreases monotonically from $1$ at $\theta=0$ to its minimum at $\theta=\pi/2$, with $\widetilde{M}_J(\pi/2) = [(\sqrt{1+\chi/\eta}-1)/(\sqrt{1+2\chi/\eta}-1)]^{3/2} < 1$. 
	Fig.~\ref{fig:mass_angular} displays $\widetilde{M}_J(\theta)$ in polar coordinates. The radial coordinate represents $\widetilde{M}_J(\theta)$; the dashed circle marks the isotropic Newtonian limit (normalized to $1$). The plot reveals a pronounced anisotropy: the Jeans mass is largest along the direction parallel to the background field ($\theta=0$ and $\pi$) and smallest in the perpendicular directions ($\theta=\pi/2$ and $3\pi/2$). The anisotropy factor $M_J(0)/M_J(\pi/2)\approx 2.5$ for these parameters, underscoring a strong directional dependence of the collapse scale in the MOND regime.
	
	The angular dependence of the Jeans mass has important implications for structure formation in self-gravitating systems under MOND. Unlike Newtonian gravity, where collapse occurs isotropically, MOND predicts that gravitational instability is more efficient in directions perpendicular to the background field, favoring the formation of anisotropic structures. This direction-dependent collapse scale could leave observable imprints. Specifically:
	\begin{itemize}
		\item \textbf{Gravitational lensing}: Non-spherical mass distributions produce asymmetric lensing signals. Future wide-field surveys such as LSST (Vera C. Rubin Observatory) and Euclid could detect such anisotropies if a MOND-regulated Bose star or dark matter halo exists along the line of sight. Moreover, detailed studies of lensing patterns and light rings can further differentiate such objects from their general relativistic counterparts~\cite{huang2025lensing}.
		\item \textbf{Very-long-baseline interferometry (VLBI)}: Recent advances have enabled imaging of supermassive compact object shadows (e.g., Sagittarius A* by the Event Horizon Telescope). If a Bose star is present at a galactic center, its anisotropic density profile—predicted by the direction-dependent Jeans mass—could manifest as asymmetric bright rings or polarization patterns, distinguishing it from both Schwarzschild black holes and Newtonian Bose stars~\cite{li2026observational}.
		\item \textbf{Gravitational waves}: Non-spherical Bose stars in binary systems would undergo tidally induced deformations, producing unique gravitational wave signatures detectable by LISA or next-generation ground-based detectors.
	\end{itemize}
	While direct observational confirmation of these predictions is beyond the scope of this work, they provide concrete targets for future astrophysical searches. The anisotropic Jeans instability thus serves as a distinctive hallmark of MOND that could be tested with next-generation facilities.
	
	\section{Summary and outlook}\label{sec:conclusion}
	We have derived the dispersion relation of a self-gravitating Bose-Einstein condensate in the MOND framework. 
	The excitation spectrum exhibits a clear anisotropy, depending on the angle between the perturbation wavevector and the background gravitational field. 
	This anisotropy stems from the nonlinearity of the MOND Poisson equation and disappears in the Newtonian limit. 
	In the deep-MOND regime, the Jeans instability becomes direction-dependent, with perpendicular perturbations being more unstable than parallel ones. The anisotropic excitation spectrum offers a distinct signature that could distinguish MOND-like effects from standard Newtonian behavior. Although derived here for a Bose gas, the same anisotropy should appear in other classical self-gravitating systems (e.g., collisionless fluids or warm dark matter), because its origin lies in the modified gravitational response. This universality makes the effect a promising target for astrophysical observations.
	
	Future work should extend our analysis to fully general relativistic MOND frameworks and incorporate realistic density profiles and accretion models to generate synthetic images for direct comparison with VLBI observations. Additionally, numerical simulations of structure formation under MOND could test whether the anisotropic Jeans instability leads to observable filamentary or disk-like structures in the cosmic web. Such studies would provide further avenues to discriminate between MOND and particle dark matter models using next-generation astronomical facilities.
	
	\paragraph*{Acknowledgments}
	This work was supported by the Open Fund of Key Laboratory of Multiscale Spin Physics (Ministry of Education), Beijing Normal University (Grant No. SPIN2024N03), and the Scientific Research Startup Foundation for High-Level Talents at Anqing Normal University (Grant No. 241042).
	

\end{document}